\documentclass[twocolumn,preprintnumbers,superscriptaddress,amsmath,amssymb]{revtex4}

\usepackage{graphicx}
\usepackage{dcolumn}
\usepackage{bm}

\usepackage{setspace}

\begin{document}

\begin{center}
Published in Rev. Sci. Instr. 80, 043301 (2009)
\end{center}

\title{Fluorescence decay-time constants in organic liquid scintillators}

\author{T. Marrod\'an Undagoitia} 
\altaffiliation{Corresponding author. Fax: +41 44 6355704,
Telephone Number: +41 44 6356686.} 
\email{marrodan@physik.uzh.ch}
\affiliation{Physik-Department, Technische Universit\"at M\"unchen,\\         
James-Franck-Str., 85748 Garching, Germany}   
\affiliation{Physik-Institut, Universit\"at Z\"urich,\\
Winterthurerstr. 190, 8057 Z\"urich, Switzerland}
\author{F. von Feilitzsch}
\author{L. Oberauer}
\author{W. Potzel} 
\author{A. Ulrich}
\author{J. Winter} 
\author{M. Wurm} 
\affiliation{Physik-Department, Technische Universit\"at M\"unchen,\\
James-Franck-Str., 85748 Garching, Germany}

\begin{abstract}

The fluorescence decay-time constants have been measured for several
scintillator mixtures based on phenyl-o-xylylethane (PXE) and linear 
alkylbenzene (LAB) solvents. The resulting
values are of relevance for the physics performance of the proposed
large-volume liquid scintillator detector LENA (Low Energy Neutrino 
Astronomy). In particular, the
impact of the measured values to the search for proton decay via $p
\to K^+ \overline{\nu}$ is evaluated in this work.
\end{abstract}

\maketitle


\section{Introduction}

Charged particles crossing a liquid-scintillator medium deposit their
energy leading partly to fluorescence light. The lifetimes of the
molecular excited states in the medium determine the pulse shape of
the events in a detector application. For this reason, experiments
relying on pulse shape analysis for background discrimination are
always dependent on the fluorescence decay-time constants. A common
application of pulse shape analysis in neutrino experiments is the
separation of $\alpha$-events produced by natural radioactivity from
neutrino-induced
$\beta$-events\,\cite{Arpesella:2008mt}. The
pulse shape also plays an important role in the search for proton
decay via $p \to K^+ \overline{\nu}$ in the proposed large
liquid-scintillator detector LENA as the signature relies on a
double-peak structure in time\,\cite{Undagoitia:2005uu}.  The LENA
(Low Energy Neutrino Astronomy)
detector\,\cite{MarrodanUndagoitia:2006re} is planned as a
large-volume (50\,kt) observatory based on the liquid-scintillator
technology which will be highly suitable for the investigation of a
variety of topics in astrophysics, geophysics and particle physics.
In the analysis of the proton decay reaction $p \to K^+
\overline{\nu}$, the two-peak time structure is used to discriminate
proton decay events from the background due to atmospheric neutrino
events\,\cite{Undagoitia:2005uu}. The detection efficiency depends
upon the fluorescence decay-time constants and therefore, the
laboratory experiments presented here have been performed to determine
these constants for several scintillator mixtures.

Fluorescence decay measurements of liquid scintillators have been
performed earlier in feasibility studies for the Borexino experiment
with a scintillator mixture based on PC (pseudocumene) as
solvent\,\cite{Ranucci94}\cite{Elisei:1997tw}. In addition, similar
measurements have been carried out for indium-loaded scintillators for
the proposed solar neutrino experiment LENS\,\cite{FHartman07}. A
desirable scintillator for the proposed LENA detector would feature
fast decay times, large attenuation lengths and reasonable costs. This
paper reports results obtained for the fluorescence decay constants of
important current scintillator candidates. Non-metal loaded
scintillators based on PXE and LAB have been tested. In addition, the
impact of the decay constants on the sensitivity of the LENA detector
to proton decay is shown.

\section{Experimental method}

The main goal of the experiment is the determination of the
'probability density function' (PDF) which describes the
time-dependence of the photon emission in the fluorescence processes.
A sampling of the photon-emission time after excitation of the medium
is required. The experiment uses the 'start-stop'
method\,\cite{Ranucci94} which measures the emission times for single
photons. In this method, the start signal accounts for the starting
point of the pulse: approximately the time of the energy
deposition. If a light detector is able to measure a large number of
photons of an event, the start time can be extracted from the first
detected photon. A second detector monitoring the same event provides
the stop signal. This detector should be designed in such a way that
the probability of detecting a single photon of an event is in the few
\% regime. This guarantees that the probability to detect more than
one photon in an event is almost zero. The PDF of a certain
scintillator mixture results from the time difference between the
start and the stop signals.

\begin{figure}[h] 
  \begin{center}
   \includegraphics[angle=-90,width=0.38\textwidth]{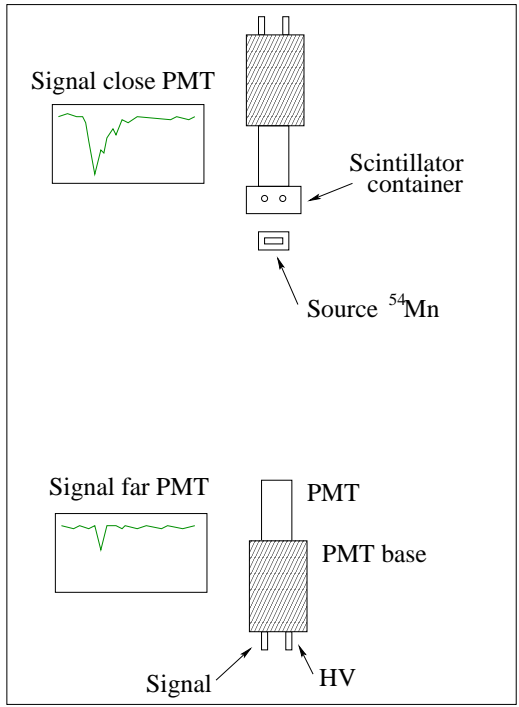}
   \caption[Setup diagram and picture]{Schematic diagram of the setup
    for the fluorescence-time measurement. Two photomultiplier tubes
    (PMTs) collect the scintillation light emitted by a scintillator
    mixture. One PMT is directly coupled to the container and the
    second is located at a distance of 30\,cm. The scintillator is
    excited by a $^{54}$Mn $\gamma$-source.\label{SetUpDiagr}}
  \end{center}	
\end{figure}

The main setup consists of two photomultipliers (PMTs) which detect
the light emitted from a liquid-scintillator sample. The sample is
irradiated by a $^{54}$Mn $\gamma$-source
(834\,keV). Figure\,\ref{SetUpDiagr} shows a scheme of the
experimental setup. A sample of scintillator is filled into a
$\sim$\,10\,ml cylindrical container made of black PTFE
(Polytetrafluoroethylene, high density) material with quartz windows
at both faces. The maximum energy for the recoiling Compton electron
for this source is 639\,keV (back-scattering, i.e. Compton-edge
energy). Compton electrons deposit their energy in the scintillator
leading to the fluorescence light-emission. One of the PMTs is
directly coupled to the scintillator container (close PMT). It detects
a large number of photons per event and thereby sets the
starting point for the event. The second PMT is placed $\sim$\,30\,cm
away from the sample (far PMT). This distance was chosen to assure
single-photon events due to the small solid angle: The probability to
detect a photon from a Compton event in the far PMT is $< 3\%$. This
value has been calculated conservatively assuming a light emission of
10\,000 photons/MeV\,\cite{Alimonti:2008gc} and an energy deposition
corresponding to the Compton edge of $^{54}$Mn.

The whole setup is located inside a light-tight box.  A NIM-based
electronic setup has been used for signal processing.  First, the
signals of both photomultipliers are amplified and directly connected
to the data acquisition system. The amplifier output for the close-PMT
signal is, in addition, connected to a discriminator and used as
trigger for the data acquisition system.

For data readout and recording, an Acqiris (Acqiris/Agilent
Technologies\,\cite{acqiris}) data acquisition system is used. In the
setup, the acquisition system is used at a sampling rate of 4\,GS/s
(giga-samples per second) together with a 10\,bit dynamic range. The
system is operated in 'external trigger mode' with the close PMT
providing the hardware trigger-signal. A Labview-based software is
used to generate a software trigger and to store the data.  When a
trigger signal from the close multiplier appears, a coincidence window
of 500\,ns is opened. If the software detects a signal above threshold
in the far PMT, both photomultiplier data pulses are stored. The
resulting files contain 50\,000~pulses covering a measuring time of
$\sim 8$\,h.

The time resolution of the system has been measured by a slight
modification of the setup just described. The $^{54}$Mn source is
replaced by a $^{22}$Na $\beta^{+}$ source which produces two 511\,keV
$\gamma$-rays from $e^+\, e^-$ annihilation. The far PMT is optically
decoupled from the fluorescence system by a black blocking filter. A
Cherenkov radiator (Plexiglas piece) is placed in front of the
far PMT. This provides an instantaneous signal as the Cherenkov
photons are produced within picoseconds time after the energy
deposition of a 511\,keV gamma. Pairs of $\gamma$-rays from the
annihilation are emitted isotropically but always in opposite
directions. Sometimes both gammas interact simultaneously in the
liquid scintillator and in the Cherenkov radiator and those coincident
events have been recorded. Using the time difference between one
511\,keV signal in the Cherenkov radiator and a second one in the
liquid scintillator sample, the time resolution of the system is
determined.

\section{Time resolution and background sources}\label{Resol&BG}

The time resolution of the fluorescence-measurement system mainly
depends upon the single-photoelectron transit \emph{time jitter} of
the PMT. Photoelectrons produced at different positions of the cathode
have different propagation distances to the first dynode resulting in
a smearing of the transit time through the tube. The specification for
the time jitter of the PMTs used is 0.98\,ns\,\cite{ETL}. Smaller
contributions to the loss in time-resolution are the error of the
determination of the start time of a pulse and the time jitter of the
electronic readout chain, the latter, however, usually being a small
contribution.

\begin{figure}[h]
  \begin{center}
    \includegraphics[width=0.49\textwidth]{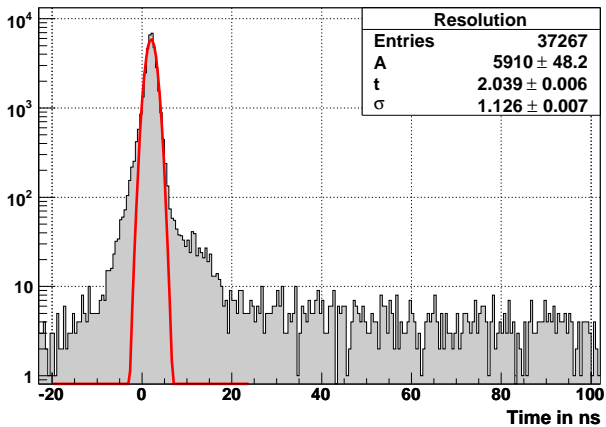}
    \caption[Results of the resolution of the system]{Distribution of
    the time differences between the start signal in the scintillator
    and the arrival of Cherenkov photons. The time resolution of the
    system (the events are plotted on a semi-logarithmic scale) is
    well described by a Gaussian curve.\label{CherResults}}
  \end{center}	
\end{figure}

Figure\,\ref{CherResults} shows the distribution of the time
differences between the start signal given by the scintillator and the
arrival of the Cherenkov photons measured in the setup with the
$^{22}$Na source.
Although the main peak is well described by a Gaussian function, there
are small contributions at times below and above the main peak. The
shoulder on the left side of the Gaussian curve is due to pre-pulses
of the single-photon tube. Further to the left, there is a
contribution which can be interpreted as photons which cross the
cathode producing a photoelectron directly in the
dynode-chain. The total contribution of the
events to the left side of the Gaussian peak is about 5\% of all
pulses.

The pulses on the right side of the Gaussian peak ($t<20$\,ns) can be
explained by the reflection of photoelectrons by the first dynode back
to the cathode. This effect occurs in about 2\% of all pulses and has
also been reported in\,\cite{Ranucci94}. The pulses at still later
times ($ t>20$\,ns) contribute about 2\% to the total number of
pulses. Their origin is unknown but may be interpreted as a weak
fluorescence of the photomultiplier housing and window as the
radioactive source was placed very close to the tube. When removing
the Plexiglas piece, it was found that the number of pulses with
$t>20$\,ns increase relative to the previous measurement pointing to a
background related to direct $\gamma$-hits on the PMT. Altogether,
about 91\% of the pulses are beneath the Gaussian curve.

In addition to the loss of time resolution due to the effects
described above random coincidences due to the dark-count rate of the
far PMT contribute to the background of the experiment. This
coincidence rate has been estimated by measuring the dark current of
the far PMT and taking the trigger rate by the close PMT. About
30~random coincidences are expected for an 8~hours measurement ($\sim
0.06\%$ of all recorded pulses).

\section{Measured samples}

The samples investigated consist of a mixture of an organic liquid
solvent and one or two wavelength-shifters as solutes. The solvents
PXE (phenyl-o-xylylethane) and LAB (linear alkylbenzene) have been
studied because of their high light yield and low risk properties.

\begin{figure}
  \begin{center}
    \includegraphics[width=0.49\textwidth]{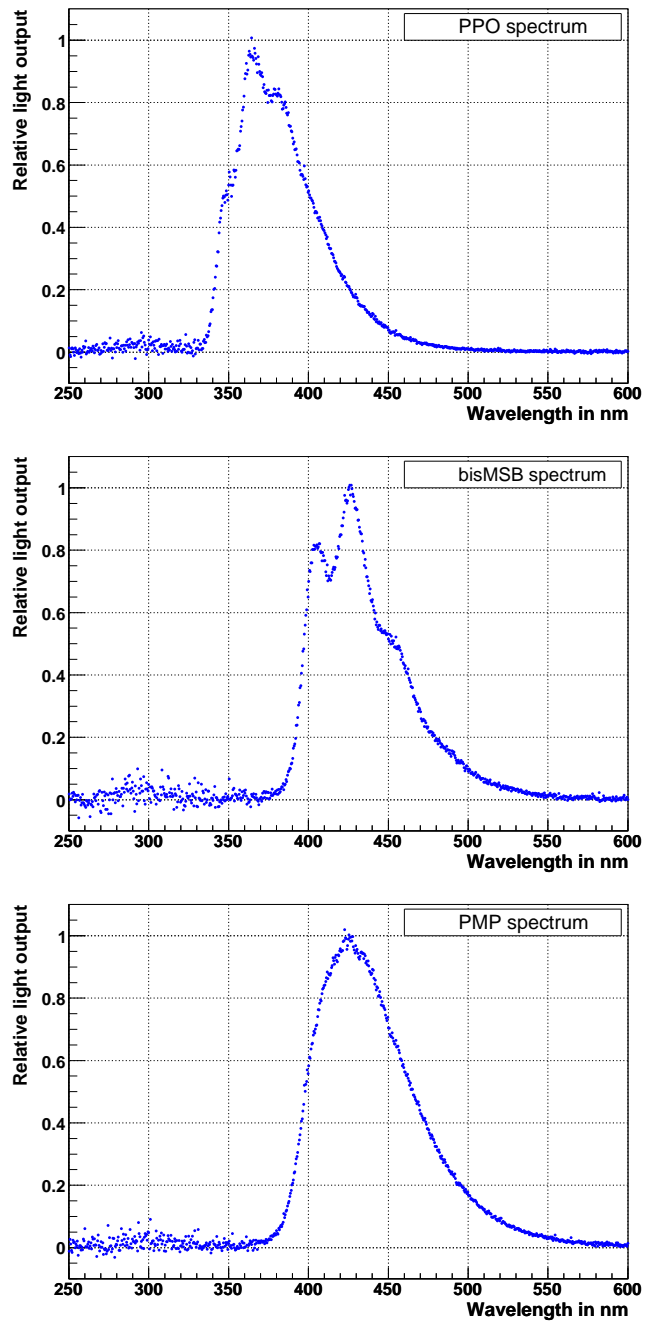}
    \caption{Emission spectra of PPO, bisMSB and PMP, all with a
    concentration of 2\,g/$\ell$ in PXE solvent.\label{Spectra}}
  \end{center}	
\end{figure}

The wavelength-shifters dissolved were PPO (2,5-diphenyl\--oxazole),
pTP (para\--terphenyl), bisMSB (1,4-bis-(o-methyl\--styryl)\--benzene)
and PMP (1-phenyl-3\--mesityl-2\--pyrazoline). To characterize these
compounds a spectrum of the emitted light has been recorded when the
compounds were dissolved in PXE. To this end, the medium was excited
with UV-light of a deuterium lamp\,\cite{PhD_TMarrodan}. The spectra
were recorded with a compact spectrometer with 0.53~nm resolution and
fiber optic coupling (Ocean Optics HR2000)\,\cite{OceanOp_Web}. They were
corrected with the spectrometer's sensitivity curve and also for
possible shifts in the
wavelength\,\cite{MarrodanSpec08}. Figure\,\ref{Spectra}
shows the emission spectra of the wavelength-shifters PPO, bisMSB and
PMP.
 The data is normalized such that the maximum of the spectrum is at
 1. All solutes have been dissolved with a concentration of 2\,g/$\ell$
 in PXE.

\section{Analysis and results}

The first step in the analysis is the determination of the start time
for both recorded pulses (close and far PMTs). The start time is
derived in a constant-fraction method by setting the starting point of
each pulse to the 40\% height in the leading edge of the pulse.  The
distribution of the photon emission time results from the difference
between the starting point of the close PMT pulse (zero-point) and
that of the far PMT. Pulse height cuts are performed to reject
dark-current pulses and overflow pulses below and above the single
photon peak, respectively. In addition, pulses with more than one peak
in the single-photon PMT are also rejected (double pulses or noise
pulses).

The probability density function (PDF), $F(t)$, of the
scintillation-light emission can be described by a convolution of a
multi-exponential function with the time resolution of the measuring
system:
\begin{equation}\label{FitFormula}
F(t)=\left( \sum_{i} \frac{N_i}{\tau_i}\cdot
e^{-\frac{(t-t_0)}{\tau_i}} \right) \otimes R(t)
\end{equation}
where $\tau_i$ is the decay-time constant of the exponential decay
function $i$ and $N_i$ is the contribution of this component such that
$\sum_i N_i = 1$; $t_0$ is the time at which the exponential decay
starts and $R(t)$ is the time-resolution function.  In good
approximation, the time resolution can be represented by a Gaussian
curve:
\begin{equation}
 R(t)= e^{-\frac{t^2}{2 \sigma^2}}
\end{equation}\label{ResGauss}
where $\sigma$ is a known parameter.  The time resolution of the
system is affected by the uncertainty in the determination of the
pulse start-time mentioned above. For the close PMT, the start time is
more accurate for scintillators with a high light yield due to the
better photon statistics. Therefore, the resolution depends on the
scintillator sample and has to be determined for every mixture. This
has been achieved by fitting a Gaussian curve to the left (sharply
rising) part of the measured PDF distribution. This method gives
compatible results with the direct time-resolution measurement
described in section\,\ref{Resol&BG}.

\begin{figure*}[t]
  \begin{center}
    \includegraphics[width=0.99\textwidth]{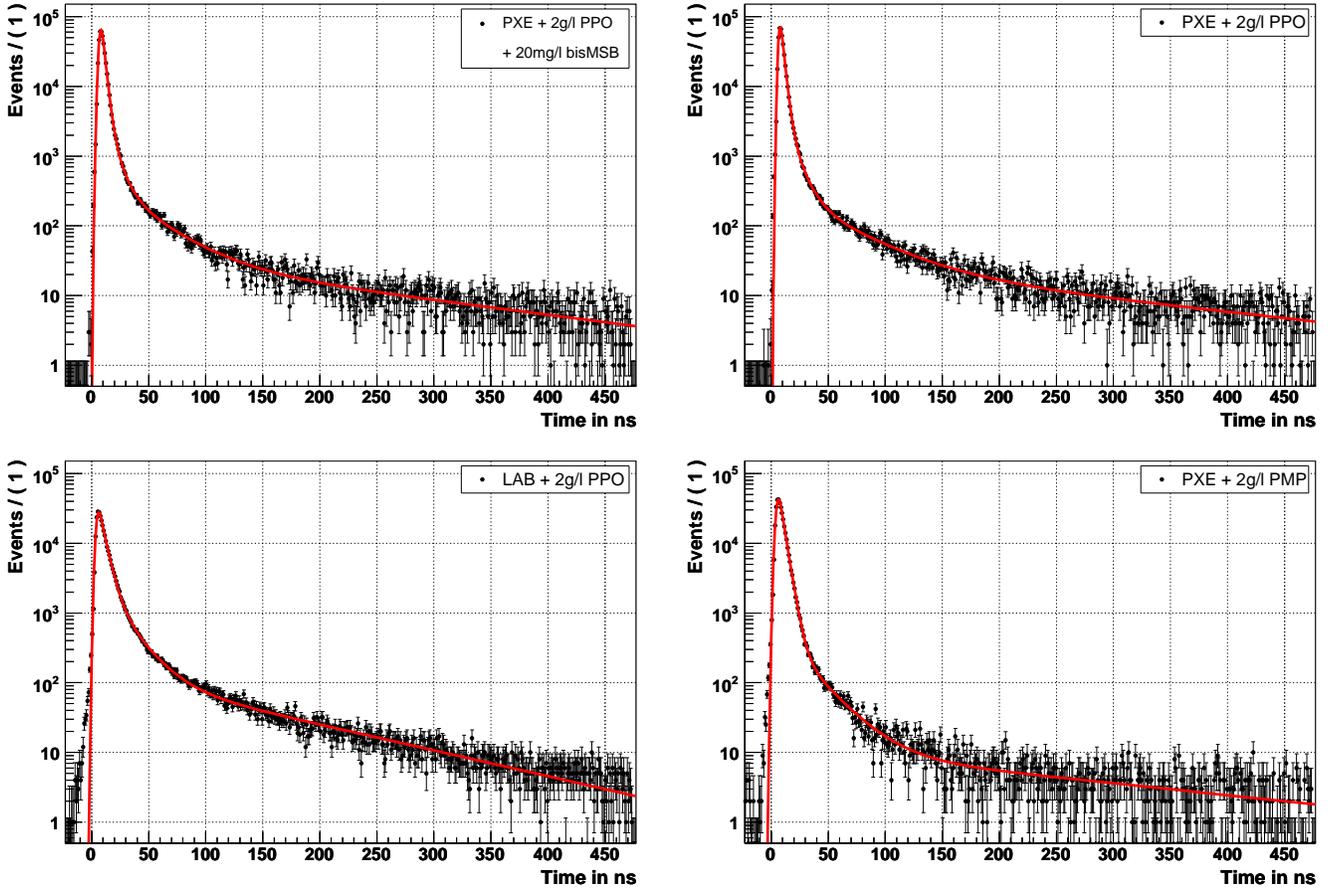}

    \caption{Probability density function of the scintillator mixtures
      (mix) PXE + 2\,g/$\ell$\,PPO + 20\,mg/$\ell$\,bisMSB, PXE +
      2\,g/$\ell$\,PPO, LAB + 2\,g/$\ell$\,PPO and PXE +
      2\,g/$\ell$\,PMP.  The plots show the time dependence of the
      light emission fitted (curve in red) according to
      equation\,\ref{FitFormula}.\label{PDFs}}
  \end{center}	
\end{figure*}

The ROOT-based Roofit\,\cite{Roofit} toolkit is used to perform the
fit of equation\,\ref{FitFormula} to the measured data.  In its
default configuration, Roofit performs a maximum likelihood fit and
minimizes $-\log (L)$ where $L$ is the likelihood function. This
minimization is performed by calling the Minuit\,\cite{James:1975dr}
package from CERN libraries.  Figure\,\ref{PDFs} shows the measured
PDFs for four examples of different scintillator mixtures: PXE +
2\,g/$\ell$\,PPO + 20\,mg/$\ell$\,bisMSB, PXE + 2\,g/$\ell$\,PPO, LAB
+ 2\,g/$\ell$\,PPO and PXE + 2\,g/$\ell$\,PMP.
The data is presented in histograms with a bin size of 1\,ns together
with the statistical errors. A fit using the convolution of the sum of
several exponential functions with a Gaussian curve is also depicted
by the red curve.

The abbreviated names of the mixtures for which the fluorescence
decay-time constants have been measured are summarized in
table\,\ref{Mixt_name}.
\begin{table}
  \caption{Nomenclature used for the mixtures of which the
      fluorescence decay constants have been measured. The sub-index
      $c$ corresponds to the concentration of the corresponding
      solute.\label{Mixt_name}}
  \vspace{0.3cm}
  \begin{tabular}{|c|c|}
    \hline Name & Mixture\\
    \hline \hline PP$c$ & PXE + $c$\,g/$\ell$\,PPO\\
    \hline LP$c$ & LAB + $c$\,g/$\ell$\,PPO\\
    \hline PM$c$ & PXE + $c$\,g/$\ell$\,PMP\\
    \hline PPb & PXE+2\,g/$\ell$\,PPO+20\,mg/$\ell$\,bisMSB\\
    \hline PTb & PXE+2\,g/$\ell$\,pTP+20\,mg/$\ell$\,bisMSB\\
    \hline
  \end{tabular}
\end{table}
 The sub-index $c$ corresponds to the concentration of the
corresponding solute.

Tables\,\ref{PDF_Tables} - \ref{ResultsPMi} summarize the parameters
of the fits according to equation\,\ref{FitFormula}. Depending on the
scintillator mixture and the solute concentration, either three or
four decay components perform the best fit. Fit parameters are the
time resolution $\sigma$, the decay constants $\tau_i$, the amplitudes
$N_i$ of each exponential function and the start time $t_0$ of the
decay function.  The amplitude of the last component ($N_{l,cal}$,
third of fourth depending on the fit) has been calculated according to
$\sum_i N_i = 1$ using the fit values. Its error depends on the other
amplitudes. To estimate a maximum error ($\Delta N_{max}$) for this
parameter, quadratic error propagation has been used. The
uncertainties $\Delta \tau_i$ and $\Delta N_i$ consist of a
statistical error $\delta_{stat}$ which is obtained from the fit and a
systematic error, $\delta_{sys}$. The latter has been derived from a
systematic study where the parameter $\sigma$ is varied within one
standard deviation of its value ($\sigma+\Delta\sigma$, 
$\sigma-\Delta\sigma$). It was found that the systematic error is of 
the same order of  magnitude as the statistic one. Both systematic
and statistic uncertainties are in general asymmetric.
\begin{table*}[t]
  \caption{Results of the fits for the mixtures based on PXE +
    2\,g/$\ell$ PPO and 2\,g/$\ell$ pTP, both with 20\,mg/$\ell$
    bisMSB as secondary wavelength-shifter. All decay times $\tau_i$,
    the time $t_0$, as well as the time resolution $\sigma$ are given
    in ns. The amplitudes $N_i$ are given in \% with $N_{l,cal}$
    calculated from $\sum_{i} N_i = 1$. The errors $\Delta \tau_i$ and
    $\Delta N_i$ include both statistic and systematic
    uncertainties. \label{PDF_Tables}}
  \begin{center}
    \begin{tabular}{|c||c|c|}
      \hline
      Mixture (mix) & PPb & PTb \\
      \hline 
      \hline 
      $\sigma$ (ns) & $1.61\pm 0.01$ & $1.50\pm 0.01$\\
      \hline 
    \end{tabular}
    \vspace{0.1cm}
  
    \begin{tabular}{|c||c|c|c|c|c|}
      \hline
      mix & $\tau_1\pm \Delta \tau_1$ 
      & $\tau_2 \pm \Delta \tau_2$ 
      & $\tau_3 \pm \Delta \tau_3$
      & $\tau_4 \pm \Delta \tau_4$ 
      & $t_0 \pm \Delta t_0$ \\
      \hline 
      \hline 
      PPb & $2.61_{-0.02}^{+0.02}{}_{-0.03}^{+0.02}$ 
      & $7.7_{-0.3}^{+0.2}{}_{-0.3}^{+0.2}$
      & $34_{-2}^{+2}{}_{-1}^{+0}$ &$205_{-11}^{+13}{}_{-0}^{+1}$
      &  $6.753_{-0.006}^{+0.006}{}_{-0.01}^{+0.01}$\\
      \hline 
      PTb & $2.38_{-0.04}^{+0.04}{}_{-0.04}^{+0.05}$  
      & $7.3_{-0.2}^{+0.2}{}_{-0.1}^{+0.1}$  
      & $35_{-2}^{+2}{}_{-0}^{+1}$ &$162_{-9}^{+10}{}_{-2}^{+1} $
      &  $6.664_{-0.008}^{+0.008}{}_{-0.02}^{+0.02}$ \\
      \hline 
    \end{tabular}
    \vspace{0.1cm}

    \begin{tabular}{|c||c|c|c|c|}
      \hline
      mix & $N_1\pm \Delta N_1$ 
      & $N_2\pm \Delta N_2$ 
      & $N_3\pm \Delta N_3$  
      & $N_{l,cal} \pm \Delta N_{max}$ \\
      \hline 
      \hline
      PPb & $0.853_{-0.008}^{+0.007}{}_{-0.006}^{+0.005}$  
      & $0.097_{-0.006}^{+0.007}{}_{-0.005}^{+0.005}$  & 
      $0.034_{-0.002}^{+0.002}{}_{-0}^{+0.001}$ 
      &  $0.02\pm 0.01$ \\
      \hline 
      PTb & $0.62_{-0.01}^{+0.01}{}_{-0.01}^{+0.03}$  
      & $0.30_{-0.01}^{+0.01}{}_{-0.01}^{+0}$  
      & $0.057_{-0.002}^{+0.002}{}_{-0}^{+0.003}$ 
      &  $0.02\pm 0.01$ \\
      \hline 
    \end{tabular}
  \end{center}	
\end{table*}
To evaluate the quality of a certain fit, both the convergence of the
fit and the value of the likelihood function ($-\log (L)$) have been
considered\,\cite{PhD_TMarrodan}.

\begin{table*}
  \caption{Results of the fits for mixtures of the solvent PXE with
    different concentrations of PPO. Further details as in
    table\,\ref{PDF_Tables}.\label{ResultsPPi}}
  \begin{center}
    \begin{tabular}{|c||c|c|c|c|}
      \hline  
      mix & PP1 & PP2 & PP6 &  PP10 \\
      \hline 
      \hline
      $\sigma$ (ns) & $1.45\pm 0.01$ 
      & $1.229\pm 0.007$& $1.066\pm 0.006$ & $1.171\pm 0.007$\\
      \hline 
    \end{tabular}
    \vspace{0.1cm}
  
    \begin{tabular}{|c||c|c|c|c|c|}
      \hline
      mix & $\tau_1\pm \Delta \tau_1$ 
      & $\tau_2 \pm \Delta \tau_2$ 
      & $\tau_3 \pm \Delta \tau_3$
      & $\tau_4 \pm \Delta \tau_4$ 
      & $t_0 \pm \Delta t_0$ \\
      \hline 
      \hline 
      PP1 & $3.16_{-0.05}^{+0.04}{}_{-0.06}^{+0.06}$  
      & $7.7_{-0.7}^{+0.7}{}_{-0.5}^{+0.9}$  
      & $34_{-3}^{+3}{}_{-2}^{+2}$ &$218_{-19}^{+25} {}_{-6}^{+12}$
      & $6.631_{-0.006}^{+0.006}{}_{-0.01}^{+0.01}$ \\
      \hline 
      PP2 & $2.63_{-0.02}^{+0.02}{}_{-0.02}^{+0.01}$ 
      & $8.8_{-0.3}^{+0.3}{}_{-0.2}^{+0.1}$
      & $43_{-3}^{+3}{}_{-0}^{+0}$ &$242_{-21}^{+27} {}_{-1}^{+2}$
      &  $6.661_{-0.005}^{+0.005}{}_{-0.008}^{+0.007}$ \\
      \hline 
      PP6 & $2.03_{-0.01}^{+0.01}{}_{-0.01}^{+0.01}$ 
      & $9.0_{-0.3}^{+0.3}{}_{-0.1}^{+0.1}$
      & $47_{-3}^{+3}{}_{-1}^{+0}$ &$203_{-13}^{+16} {}_{-0}^{+1}$
      &  $6.733_{-0.004}^{+0.004}{}_{-0.007}^{+0.006}$ \\
      \hline 
      PP10 & $2.04_{-0.01}^{+0.01}{}_{-0.01}^{+0.02}$ 
      & $8.6_{-0.4}^{+0.4}{}_{-0.1}^{+0.2}$
      & $40_{-2}^{+2}{}_{-3}^{+3}$ &$207_{-14}^{+14}{}_{-0}^{+1}$
      & $6.627_{-0.005}^{+0.005}{}_{-0.008}^{+0.008}$ \\
      \hline
    \end{tabular}
    \vspace{0.1cm}

    \begin{tabular}{|c||c|c|c|c|}
      \hline
      mix & $N_1\pm \Delta N_1$ 
      & $N_2\pm \Delta N_2$ 
      & $N_3\pm \Delta N_3$  
      & $N_{l,cal} \pm \Delta N_{max}$ \\
      \hline 
      \hline

      PP1 & $0.84_{-0.02}^{+0.02}{}_{-0.02}^{+0.03}$  
      & $0.12_{-0.01}^{+0.02}{}_{-0.02}^{+0.02}$  & 
      $0.029_{-0.002}^{+0.003}{}_{-0.002}^{+0.001}$ 
      & $0.011\pm 0.028$ \\
      \hline 
      PP2 & $0.852_{-0.005}^{+0.005}{}_{-0.003}^{+0.002}$  
      & $0.103_{-0.004}^{+0.005}{}_{-0.003}^{+0.002}$  & 
      $0.031_{-0.001}^{+0.001}{}_{-0}^{+0}$ 
      &  $0.014\pm 0.007$ \\
      \hline 
      PP6 & $0.822_{-0.003}^{+0.003}{}_{-0.002}^{+0.001}$  
      & $0.105_{-0.002}^{+0.002}{}_{-0.001}^{+0.001}$  & 
      $0.046_{-0.002}^{+0.002}{}_{-0}^{+0}$ 
      &  $0.027\pm 0.004$ \\
      \hline 
      PP10 & $0.802_{-0.004}^{+0.004}{}_{-0.002}^{+0.002}$  
      & $0.105_{-0.003}^{+0.003}{}_{-0.002}^{+0.001}$  & 
      $0.061_{-0.002}^{+0.002}{}_{-0.001}^{+0}$ 
      & $0.032\pm 0.005$ \\
      \hline
    \end{tabular}
  \end{center}	
\end{table*}

\begin{table*}
  \caption{Results of the fits for mixtures of the solvent LAB with
    different concentrations of PPO. Further details as in
    table\,\ref{PDF_Tables}.\label{ResultsLPi}}
  \begin{center}
    \begin{tabular}{|c||c|c|c|c|}
      \hline  
      mix & LP1 & LP2 & LP6 & LP10 \\
      \hline 
      \hline
      $\sigma$ (ns) & $1.92\pm 0.01$ 
      & $1.58\pm 0.01$& $1.236\pm 0.009$ & $1.133\pm 0.008$\\
      \hline 
    \end{tabular}
    \vspace{0.1cm}
  
    \begin{tabular}{|c||c|c|c|c|c|}
      \hline
      mix & $\tau_1\pm \Delta \tau_1$ 
      & $\tau_2 \pm \Delta \tau_2$ 
      & $\tau_3 \pm \Delta \tau_3$
      & $\tau_4 \pm \Delta \tau_4$ 
      & $t_0 \pm \Delta t_0$ \\
      \hline 
      \hline 
      LP1 & $7.46_{-0.07}^{+0} {}_{-0.03}^{+0.04}$  
      & $22.3_{-0.6}^{+0.6}{}_{-0.2}^{+0.2}$ 
      &$115_{-3}^{+4} {}_{-1}^{+0}$ & -
      & $4.01_{-0.01}^{+0.01}{}_{-0.01}^{+0.01}$ \\
      \hline 
      LP2  & $5.21_{-0.05}^{+0.04}{}_{-0.03}^{+0.03}$ 
      & $18.4_{-0.6}^{+0.6}{}_{-0.2}^{+0.2}$
      & $118_{-2}^{+3} {}_{-1}^{+0}$ & -
      & $4.161_{-0.009}^{+0.009}{}_{-0.01}^{+0.009}$ \\
      \hline 
       LP6  & $2.71_{-0.08}^{+0.06}{}_{-0.08}^{+0.07}$ 
      & $6.7_{-0.6}^{+0.6}{}_{-0.4}^{+0.5}$
      & $30_{-2}^{+2}{}_{-1}^{+1}$ & $136_{-5}^{+6}{}_{-2}^{+2}$
      &  $4.577_{-0.007}^{+0.007}{}_{-0.01}^{+0.01}$ \\
      \hline 
       LP10  & $1.94_{-0.05}^{+0.04}{}_{-0.04}^{+0.04}$ 
      & $5.9_{-0.3}^{+0.3}{}_{-0.2}^{+0.2}$
      & $26.9_{-0.9}^{+1}{}_{-0.3}^{+1}$ & $137_{-4}^{+5}{}_{-1}^{+0}$
      &  $4.550_{-0.007}^{+0.007}{}_{-0.01}^{+0.01}$ \\
      \hline
    \end{tabular}
    \vspace{0.1cm}

    \begin{tabular}{|c||c|c|c|c|}
      \hline
      mix & $N_1\pm \Delta N_1$ 
      & $N_2\pm \Delta N_2$ 
      & $N_3\pm \Delta N_3$  
      & $N_{l,cal} \pm \Delta N_{max}$ \\
      \hline 
      \hline
      LP1  & $0.759_{-0.009}^{+0.009}{}_{-0.003}^{+0.004}$  
      & $0.21_{-0.008}^{+0.008}{}_{-0.003}^{+0.003}$  & 
      - &  $0.031\pm 0.012$ \\
      \hline 
      LP2  & $0.777_{-0.007}^{+0.007}{}_{-0.003}^{+0.003}$  
      & $0.170_{-0.006}^{+0.006}{}_{-0.003}^{+0.002}$  & 
      - &  $0.053\pm 0.009$ \\
      \hline 
      LP6  & $0.65_{-0.04}^{+0.03}{}_{-0.03}^{+0.03}$  
      & $0.21_{-0}^{+0.02}{}_{-0.02}^{+0.03}$  & 
      $0.100_{-0.004}^{+0.004}{}_{-0.003}^{+0.002}$
      &  $0.040\pm 0.036$ \\
      \hline 
      LP10  & $0.56_{-0.02}^{+0.02}{}_{-0.01}^{+0.01}$  
      & $0.27_{-0.01}^{+0.01}{}_{-0.01}^{+0.01}$  & 
      $0.133_{-0.004}^{+0.004}{}_{-0.02}^{+0.01}$
      &  $0.037\pm 0.023$ \\
      \hline
    \end{tabular}
  \end{center}	
\end{table*}

\begin{table*}
  \caption{Results of the fits for mixtures of the solvent PXE with
    different concentrations of PMP. Further details as in
    table\,\ref{PDF_Tables}.\label{ResultsPMi}}
  \begin{center}
    \begin{tabular}{|c||c|c|c|c|}
      \hline  
      mix & PM1 & PM2 & PM6 &  PM10 \\
      \hline 
      \hline
      $\sigma$ (ns) & $2.12\pm 0.01$ 
      & $1.77\pm 0.01$& $1.362\pm 0.009$ & $1.258\pm 0.009$\\
      \hline 
    \end{tabular}
    \vspace{0.1cm}
  
    \begin{tabular}{|c||c|c|c|c|}
      \hline
      mix & $\tau_1\pm \Delta \tau_1$ 
      & $\tau_2 \pm \Delta \tau_2$ 
      & $\tau_3 \pm \Delta \tau_3$ 
      & $t_0 \pm \Delta t_0$ \\
      \hline 
      \hline 
      PM1 & $4.30_{-0.01}^{+0.01}{}_{-0.02}^{+0.01}$  
      & $21.4_{-0.6}^{+0.6}{}_{-0.2}^{+0.2}$  
      & $212_{-18}^{+22}{}_{-1}^{+3}$ 
      &  $4.466_{-0.007}^{+0.007}{}_{-0.009}^{+0.009}$ \\
      \hline 
      PM2 & $4.15_{-0.01}^{+0.01}{}_{-0.01}^{+0.01}$ 
      & $23.7_{-0.8}^{+0.8}{}_{-0.2}^{+0.2}$
      & $255_{-20}^{+23}{}_{-3}^{+2}$ 
      &  $4.557_{-0.007}^{+0.007}{}_{-0.009}^{+0.008}$ \\
      \hline 
      PM6 & $3.62_{-0.01}^{+0.01}{}_{-0.01}^{+0.01}$ 
      & $18.5_{-0.7}^{+0.7}{}_{-0.2}^{+0.3}$
      & $137_{-5}^{+5}{}_{-1}^{+1}$ 
      &  $4.375_{-0.006}^{+0.006}{}_{-0.008}^{+0.008}$\\
      \hline 
      PM9 & $3.44_{-0.02}^{+0.02}{}_{-0.01}^{+0.01}$
      & $17.6_{-0.7}^{+0.7}{}_{-0.3}^{+0.3}$ &$135_{-4}^{+4}{}_{-1}^{+1}$
      &  $4.381_{-0.006}^{+0.006}{}_{-0.007}^{+0.008}$\\

      \hline
    \end{tabular}
    \vspace{0.1cm}

    \begin{tabular}{|c||c|c|c|}
      \hline
      mix & $N_1\pm \Delta N_1$ 
      & $N_2\pm \Delta N_2$  
      & $N_{l,cal} \pm \Delta N_{max}$ \\
      \hline 
      \hline
      PM1 & $0.956_{-0.001}^{+0.001}{}_{-0.001}^{+0}$  
      & $0.032_{-0.001}^{+0.001}{}_{-0}^{+0}$  &  $0.012\pm 0.001$ \\
      \hline 
      PM2 & $0.959_{-0.001}^{+0.001}{}_{-0.001}^{+0}$  
      & $0.035_{-0.001}^{+0.001}{}_{-0.001}^{+0}$   
      &  $0.006\pm 0.001$ \\
      \hline 
      PM6 &$0.935_{-0.002}^{+0.002}{}_{-0.001}^{+0}$  
      & $0.048_{-0.002}^{+0.002}{}_{-0}^{+0.001}$  & $0.017\pm 0.003$ \\
      \hline 
      PM9 & $0.912_{-0.006}^{+0.006}{}_{-0.001}^{+0.001}$  
      & $0.064_{-0.002}^{+0.002}{}_{-0.001}^{+0.001}$  
      &  $0.024\pm 0.006$ \\
      \hline
    \end{tabular}
  \end{center}	
\end{table*}

\section{Discussion}

For each sample, the shortest decay-time constant obtained, $\tau_1$,
has a value of few nanoseconds ($2-8$\,ns) and represents the main
contribution of the photon emission (its amplitude varies between 60\%
and 95\% of the total emission). This contribution is
related\,\cite{Birk} to the transition between the lowest excited
singlet spin-state $S_{1}$ in the solute and its ground state
$S_{0}$. The additional two or three components arise from further
molecular processes such as the de-excitation of electrons in triplet
spin-states. In general, the data shows that mixtures based on LAB
solvent show longer decay constants. This observation has been made by
other groups\,\cite{SpecLAB_Buck}\cite{CKraus_MChen} as well. Concerning the
wavelength-shifters, PPO shows shorter values for $\tau_1$ than
PMP. The addition of the secondary wavelength-shifter bisMSB to a
mixture containing PPO has only a small effect on the decay times. 

For three scintillator systems (PXE-PPO, LAB-PPO and PXE-PMP), the
evolution of the shortest decay time constant $\tau_1$ with the solute
concentration $c$ has been studied. From the variation of this
decay-time constant, information on the energy transfer in the
different systems can be obtained. The energy transfer mechanism can
be described in the following
way\,\cite{CBuck_PhD}\cite{FHartman07}. After the energy deposition by
a charged particle, the solvent molecule (PXE or LAB) rapidly reaches
the first excited state. Via collisions with neighboring solvent
molecules, the excitation energy can propagate between solvent
molecules with a rate $k_{t}$ [s$^{-1}$]. This process is also known
as \emph{energy hopping}\,\cite{FHartmanPC}. The excitation energy
moves spatially until a solute molecule is encountered to which the
energy is transferred by dipole-dipole
interaction\,\cite{Forster59}. The measured decay constant $\tau_1$ is
mainly the sum of the energy-hopping time $\tau_{t}(c)=
\frac{n(c)}{k_{t}}$ and the intrinsic lifetime of the solute
$\tau_{s}$, $\tau_1 = \tau_s + \tau_t (c)$. The number of
solvent-solvent collision processes $n$ depends on the solute
concentration $c$, the higher the concentration the lower the value
for $n$, the faster the energy transfer from a solvent to a solute
molecule.

Assuming a linear dependence of the transfer rate on the solute
concentration\,\cite{Berlman60}\cite{Berlman61} and defining an
effective hopping rate $k_h$, it follows that: $ k_{t}/ n(c) = k_h
\cdot c/ c_0$ and the measured decay constant $\tau_1$ can be written as:
\begin{equation}
\tau_1(c) = \tau_s + \frac{c_0}{k_h \cdot c}.
 \label{FitTausC}
\end{equation}
where $c_0=1$\,g/$\ell$. By fitting this formula to the experimental data, 
the intrinsic lifetime of the solute $\tau_s$ can be extracted. Moreover, the value
of $k_h$ represents the efficiency of the energy transfer via
collisions (hopping) between solvent molecules until the energy
reaches a solute molecule.

\begin{figure}
  \begin{center}

   \includegraphics[width=0.495\textwidth]{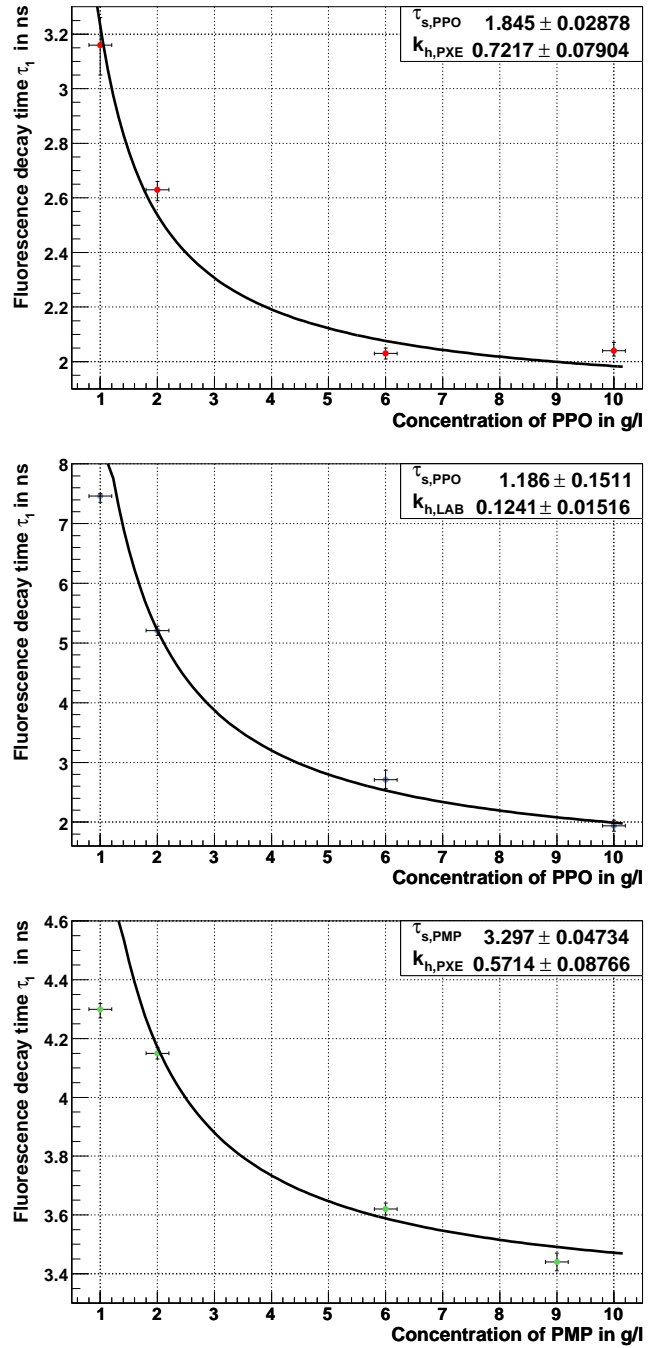}

   \caption[T1_Evol_PXEPPO]{Evolution of the fluorescence decay time
   $\tau_1$ for different concentrations of PPO in the solvents PXE
   (upper part) and LAB (middle). Bottom: evolution of $\tau_1$ for
   different concentrations of PMP in PXE.\label{T1_EvolC}}
  \end{center}	
\end{figure}

Figure\,\ref{T1_EvolC} shows the evolution of $\tau_1$ with the PPO
concentration for PPO being dissolved in the solvents PXE (upper
panel) and LAB (middle panel).
The lowest panel shows the concentration dependence for the solute PMP
dissolved in PXE.  The errors assigned to the data points are the
linear sum of the statistic and systematic uncertainties. The fit
applied to the data points corresponds to
equation\,\ref{FitTausC}. For each plot, the right upper-box shows the
fit results where the parameter $\tau_s$ is labeled with the name of
the solute and $k_h$ with the name of the solvent. The fit takes both
the uncertainties in the solute concentrations and the errors in the
decay-time constants into account, with the latter errors being
asymmetric.

From the fits, the intrinsic lifetime of the PPO singlet-state $S_1$
can be extracted.  For the PXE-PPO mixtures
$\tau_{s,\textrm{\tiny{PPO}}}= 1.85 \pm 0.03$\,ns is obtained and
$\tau_{s,\textrm{\tiny{PPO}}}= 1.19 \pm 0.15$\,ns for the mixtures of
LAB-PPO. Thus, the obtained value for $\tau_s$ is lower when the LAB
solvent is used. The solute lifetime changes with the solvent type
because the interaction between both molecules changes. In the
literature, the value for the intrinsic lifetime of the $S_{1}$ state
of PPO varies slightly. In~\cite{Berlman71} the values 1.44 and
1.8\,ns are reported. They result from different theoretical
calculations of the $S_1 \to S_0$ transition and include a factor
which takes the index of refraction of the solvent medium into account
(cyclohexane in the published cases). Finally,
in\,\cite{Alimonti:2000wj} the value 1.6\,ns has been used. The measured values for the PPO lifetime do not agree within the given error. We attribute this difference to solvent effects. The theoretical values given in the literature are in fair agreement with the results for $\tau_s$ presented here.

The parameter $k_h$ quantifies, as explained above, the transfer of
energy between solvent molecules via collisions. This transfer rate is
faster for PXE than for LAB which explains why for low PPO
concentrations the measured $\tau_1$ values are much shorter than
those for PPO dissolved in LAB. As an example, for a concentration of
2\,g/$\ell$ PPO, $\tau_{t,\textrm{\tiny{PXE}}}=2.63$\,ns and
$\tau_{t,\textrm{\tiny{LAB}}}=4.03$\,ns (see
equation\,\ref{FitTausC}).  

The obtained value for the intrinsic lifetime of PMP,
$\tau_{s,\textrm{\tiny{PMP}}} = 3.30 \pm 0.05$\,ns is close to the
measured value of 3.01\,ns given in reference\,\cite{Guesten1978}.
$\tau_{s,\textrm{\tiny{PMP}}}$ is significantly larger than the
intrinsic lifetime of PPO. This might be due to the fact that the
overlap between absorption and emission spectra in PMP is
significantly lower than in PPO. The value of
$k_{h,\textrm{\tiny{PXE}}}$ of PMP is compatible to the one measured
in the PXE-PPO system ($k_{h,\textrm{\tiny{PXE}}}=0.57\pm 0.09$ when
PMP is used and $k_{h,\textrm{\tiny{PXE}}}=0.72\pm 0.08$ for the
mixture with PPO).  The measured $\tau_1$ constants decrease with
increasing concentration and can well be fitted by
equation\,\ref{FitTausC}. However, for large detectors solute
self-absorption increases with concentration as well leading to longer
$\tau_1$. This is due to the additional time which the light reemission 
takes\,\cite{PhD_TMarrodan}.

\section{Impact on scintillation-detector performance}

The measurements of the fluorescence decay-time constants show that
the main component of the light emission ($\tau_1$, $N_1$) is
significantly dependent on the scintillator mixture.  The impact of
this variation on the potential of the proposed LENA detector to
search for proton decay ($p \to K^+ \overline{\nu}$) has been
investigated in the present paper. The signature of these decay events
consists of two pulses close in time\,\cite{Undagoitia:2005uu}.  The
first arises from the energy deposited by the kaon, the second from
the energy of its decay particles ($K^+ \to \mu^+ + \nu_{\mu}$ and
$K^+ \to \pi^+ + \pi^{0}$, with 63.4 and 20.9\% branching ratios,
respectively). Due to the short lifetime of the kaon (12.4\,ns) these
signals overlap significantly. However, the double-peak structure in
the time can be used to discriminate those events from atmospheric
neutrino background.  The sensitivity of LENA to search for proton
decay has been studied with a Geant4\,\cite{Geant}-based Monte Carlo
simulation presented in reference\,\cite{Undagoitia:2005uu}. In those
simulations, a main decay time constant $\tau_1$ of 3.5\,ns was
used. A sensitivity to the proton lifetime of $4\cdot 10^{34}$\,y has
been derived from the analysis.

Using the results of the present work, new simulations have been
performed for two possible values of $\tau_1$ (3 and 6\,ns) which are
reasonable according to the presented measurements. For some samples,
the decay time constant $\tau_1$ measured was even shorter than the
chosen values. However, light propagation also has an influence on the
pulse shape as scattering and solute self-absorption smear slightly
the time information. The values of the absorption and scattering
lengths are set to 20\,m each as laboratory measurements have shown
that those are realistic values. The resulting attenuation length is
10\,m. All other scintillation constants were kept at their previous
values. The time jitter of the photomultipliers is also included in
the simulation. Further details are discussed
in~\cite{Undagoitia:2005uu}. The resulting efficiencies for proton decay
detection in LENA via $p \to K^+ \overline{\nu}$ are 65\% and 56\% for
the decay values $\tau_1 = 3$\,ns and $\tau_1 = 6$\,ns,
respectively. This means that the choice of the scintillator mixture
can have an effect of about 20\% on the achievable proton lifetime
limit.

\section{Conclusions}

Fluorescence decay-time constants have been studied for several
scintillator mixtures which could be relevant for the LENA
detector. The effects of variation of the solvent type, solute type
and concentration have been investigated systematically. In the
measured samples, there is a short decay-time constant ($\tau_1$)
which accounts for the main part of the emitted light (60 to
95\%). For this decay constant, large differences from 2 to 8\,ns have
been obtained for various scintillator mixtures. Concerning the
fluorescence decay-time constants, attenuation lengths and the solute
self-absorption, a mixture containing PXE, about 2\,g/$\ell$ PPO and
$\sim 20$\,mg/$\ell$ bisMSB shows the best performance for LENA. The
emission spectrum of bisMSB is shifted to longer wavelength values
(see figure\,\ref{Spectra}) in comparison with spectra of usual
primary wavelength-shifters like PPO. Since absorption and scattering
lengths increase with the wavelength, the addition of bisMSB is
favored. In principle, mixtures of PXE with PPO show a shorter
$\tau_1$ but the self absorption of the wavelength-shifter plays also
a role in the light propagation over large distances smearing the time
information. The addition of a mineral oil like dodecane is currently
under discussion. It has the advantage of being very transparent (long
attenuation length) and has a large number of free protons. The latter
is important for the detection of $\overline{\nu}_e$ via their capture
on free protons\,\cite{Wurm:2007cy}. However, the addition of dodecane
would result in slower decay-time constants\,\cite{MWurm_Dipl}. 

The performance of the liquid scintillator in LENA has direct
implications on the physics potential. This has been studied within a
Monte Carlo simulation. If the shortest fluorescence decay-time
constant is increased from 3 to 6\,ns, the detection efficiency and
therefore the sensitivity is reduced by 20\%.

\section*{Acknowledgments}

We want to thank C. Buck and F. X. Hartmann for valuable
discussions. This work has been supported by funds of the
Maier-Leibnitz-Laboratorium (Munich), the Deutsche
Forschungsgemeinschaft (DFG) transregio TR27 (Neutrinos and beyond)
and the Munich Cluster of Excellence (Origin and structure of the
Universe).


\begin{thebibliography}{29}
\expandafter\ifx\csname natexlab\endcsname\relax\def\natexlab#1{#1}\fi
\expandafter\ifx\csname bibnamefont\endcsname\relax
  \def\bibnamefont#1{#1}\fi
\expandafter\ifx\csname bibfnamefont\endcsname\relax
  \def\bibfnamefont#1{#1}\fi
\expandafter\ifx\csname citenamefont\endcsname\relax
  \def\citenamefont#1{#1}\fi
\expandafter\ifx\csname url\endcsname\relax
  \def\url#1{\texttt{#1}}\fi
\expandafter\ifx\csname urlprefix\endcsname\relax\def\urlprefix{URL }\fi
\providecommand{\bibinfo}[2]{#2}
\providecommand{\eprint}[2][]{\url{#2}}

\bibitem[{\citenamefont{Arpesella
  et~al.}(2008{\natexlab{a}})}]{Arpesella:2008mt}
\bibinfo{author}{\bibfnamefont{C.}~\bibnamefont{Arpesella}}
  \bibnamefont{et~al.} (\bibinfo{collaboration}{{Borexino Collaboration}}),
  \bibinfo{journal}{Phys. Rev. Lett.} \textbf{\bibinfo{volume}{101}},
  \bibinfo{pages}{091302} (\bibinfo{year}{2008}{\natexlab{a}}),
  \eprint{arXiv:0805.3843v2 [astro-ph]}.

\bibitem[{\citenamefont{{T. Marrod\'an Undagoitia}
  et~al.}(2005)}]{Undagoitia:2005uu}
\bibinfo{author}{\bibnamefont{{T. Marrod\'an Undagoitia}}}
  \bibnamefont{et~al.}, \bibinfo{journal}{Phys. Rev. D}
  \textbf{\bibinfo{volume}{72}}, \bibinfo{pages}{075014}
  (\bibinfo{year}{2005}), \eprint{arXiv:hep-ph/0511230}.

\bibitem[{\citenamefont{{T. Marrod\'an Undagoitia}
  et~al.}(2006)}]{MarrodanUndagoitia:2006re}
\bibinfo{author}{\bibnamefont{{T. Marrod\'an Undagoitia}}}
  \bibnamefont{et~al.}, \bibinfo{journal}{Prog. Part. Nucl. Phys.}
  \textbf{\bibinfo{volume}{57}}, \bibinfo{pages}{283} (\bibinfo{year}{2006}),
  \eprint{arXiv:hep-ph/0605229}.

\bibitem[{\citenamefont{Ranucci et~al.}(1994)}]{Ranucci94}
\bibinfo{author}{\bibfnamefont{G.}~\bibnamefont{Ranucci}} \bibnamefont{et~al.},
  \bibinfo{journal}{Nucl. Instrum. Meth.} \textbf{\bibinfo{volume}{350}},
  \bibinfo{pages}{338} (\bibinfo{year}{1994}).

\bibitem[{\citenamefont{Elisei et~al.}(1997)}]{Elisei:1997tw}
\bibinfo{author}{\bibfnamefont{F.}~\bibnamefont{Elisei}} \bibnamefont{et~al.},
  \bibinfo{journal}{Nucl. Instrum. Meth.} \textbf{\bibinfo{volume}{A400}},
  \bibinfo{pages}{53} (\bibinfo{year}{1997}).

\bibitem[{\citenamefont{Buck et~al.}(2007)\citenamefont{Buck, Hartmann, Motta,
  and Schoenert}}]{FHartman07}
\bibinfo{author}{\bibfnamefont{C.}~\bibnamefont{Buck}},
  \bibinfo{author}{\bibfnamefont{F.~X.} \bibnamefont{Hartmann}},
  \bibinfo{author}{\bibfnamefont{D.}~\bibnamefont{Motta}}, \bibnamefont{and}
  \bibinfo{author}{\bibfnamefont{S.}~\bibnamefont{Schoenert}},
  \bibinfo{journal}{Chem. Phys. Lett.} \textbf{\bibinfo{volume}{435}},
  \bibinfo{pages}{252} (\bibinfo{year}{2007}).

\bibitem[{\citenamefont{Alimonti}(2008)}]{Alimonti:2008gc}
\bibinfo{author}{\bibfnamefont{G.}~\bibnamefont{Alimonti}}
  (\bibinfo{collaboration}{{Borexino Collaboration}}) (\bibinfo{year}{2008}),
  \eprint{arXiv:0806.2400v1 [physics.ins-det]}.

\bibitem[{\citenamefont{{Agilent Technologies. Acqiris}}()}]{acqiris}
\bibinfo{author}{\bibnamefont{{Agilent Technologies. Acqiris}}},
  \bibinfo{howpublished}{www.acqiris.com/}.

\bibitem[{\citenamefont{{Electron Tubes}}()}]{ETL}
\bibinfo{author}{\bibnamefont{{Electron Tubes}}},
  \bibinfo{howpublished}{www.electrontubes.com/}.

\bibitem[{\citenamefont{{T. Marrod\'an Undagoitia}}(2008)}]{PhD_TMarrodan}
\bibinfo{author}{\bibnamefont{{T. Marrod\'an Undagoitia}}}, Ph.D. thesis,
  \bibinfo{school}{Technische Universit\"at M\"unchen},
  \bibinfo{address}{Garching, Germany} (\bibinfo{year}{2008}),
  \bibinfo{note}{http://mediatum2.ub.tum.de/doc/667813/667813.pdf}.

\bibitem[{\citenamefont{{Ocean Optics}}(2007)}]{OceanOp_Web}
\bibinfo{author}{\bibnamefont{{Ocean Optics}}} (\bibinfo{year}{2007}),
  \bibinfo{note}{http://www.oceanoptics.com/}.

\bibitem[{\citenamefont{{T. Marrod\'an Undagoitia}
  et~al.}(2009)}]{MarrodanSpec08}
\bibinfo{author}{\bibnamefont{{T. Marrod\'an Undagoitia}}} \bibnamefont{et~al.},
 \bibinfo{note}{to be published}  (\bibinfo{year}{2009}).

\bibitem[{\citenamefont{Verkerke and Kirkby}()}]{Roofit}
\bibinfo{author}{\bibfnamefont{W.}~\bibnamefont{Verkerke}} \bibnamefont{and}
  \bibinfo{author}{\bibfnamefont{D.}~\bibnamefont{Kirkby}},
  \emph{\bibinfo{title}{The roofit toolkit for data modeling}},
  \bibinfo{note}{http://roofit.sourceforge.net/}.

\bibitem[{\citenamefont{James and Roos}(1975)}]{James:1975dr}
\bibinfo{author}{\bibfnamefont{F.}~\bibnamefont{James}} \bibnamefont{and}
  \bibinfo{author}{\bibfnamefont{M.}~\bibnamefont{Roos}},
  \bibinfo{journal}{Comput. Phys. Commun.} \textbf{\bibinfo{volume}{10}},
  \bibinfo{pages}{343} (\bibinfo{year}{1975}).

\bibitem[{\citenamefont{Birks}(1964)}]{Birk}
\bibinfo{author}{\bibfnamefont{J.~B.} \bibnamefont{Birks}},
  \emph{\bibinfo{title}{The Theory and Practice of Scintillation Counting}}
  (\bibinfo{publisher}{Pergamon Press}, \bibinfo{address}{London},
  \bibinfo{year}{1964}).

\bibitem[{\citenamefont{Buck}(2007)}]{SpecLAB_Buck}
\bibinfo{author}{\bibfnamefont{C.}~\bibnamefont{Buck}},
  \bibinfo{note}{private communication} (\bibinfo{year}{2007}).

\bibitem[{\citenamefont{Kraus and Chen}(2007)}]{CKraus_MChen}
\bibinfo{author}{\bibfnamefont{C.}~\bibnamefont{Kraus}} \bibnamefont{and}
  \bibinfo{author}{\bibfnamefont{M.}~\bibnamefont{Chen}},
  \bibinfo{howpublished}{private communication} (\bibinfo{year}{2007}).

\bibitem[{\citenamefont{Buck}(2004)}]{CBuck_PhD}
\bibinfo{author}{\bibfnamefont{C.}~\bibnamefont{Buck}}, Ph.D. thesis,
  \bibinfo{school}{Ruperto-Carola Universit\"at}, \bibinfo{address}{Heidelberg,
  Germany} (\bibinfo{year}{2004}).

\bibitem[{\citenamefont{Hartmann}(2008)}]{FHartmanPC}
\bibinfo{author}{\bibfnamefont{F.~X.} \bibnamefont{Hartmann}},
 \bibinfo{note}{private communication}  (\bibinfo{year}{2008}).

\bibitem[{\citenamefont{Forster}(1959)}]{Forster59}
\bibinfo{author}{\bibfnamefont{T.}~\bibnamefont{Forster}},
  \bibinfo{journal}{Discuss. Faraday Soc.} \textbf{\bibinfo{volume}{27}},
  \bibinfo{pages}{7} (\bibinfo{year}{1959}).

\bibitem[{\citenamefont{Berlman}(1960)}]{Berlman60}
\bibinfo{author}{\bibfnamefont{I.~B.} \bibnamefont{Berlman}},
  \bibinfo{journal}{J. Chem. Phys.} \textbf{\bibinfo{volume}{33}},
  \bibinfo{pages}{1124} (\bibinfo{year}{1960}).

\bibitem[{\citenamefont{Berlman}(1961)}]{Berlman61}
\bibinfo{author}{\bibfnamefont{I.~B.} \bibnamefont{Berlman}},
  \bibinfo{journal}{J. Chem. Phys.} \textbf{\bibinfo{volume}{34}},
  \bibinfo{pages}{598} (\bibinfo{year}{1961}).

\bibitem[{\citenamefont{Berlman}(1971)}]{Berlman71}
\bibinfo{author}{\bibfnamefont{I.~B.} \bibnamefont{Berlman}},
  \emph{\bibinfo{title}{Handbook of fluorescence spectra of aromatic
  molecules}} (\bibinfo{publisher}{Academic press}, \bibinfo{address}{New York
  and London}, \bibinfo{year}{1971}).

\bibitem[{\citenamefont{Alimonti et~al.}(2000)}]{Alimonti:2000wj}
\bibinfo{author}{\bibfnamefont{G.}~\bibnamefont{Alimonti}} \bibnamefont{et~al.}
  (\bibinfo{collaboration}{{Borexino Collaboration}}), \bibinfo{journal}{Nucl.
  Instrum. Meth.} \textbf{\bibinfo{volume}{A440}}, \bibinfo{pages}{360}
  (\bibinfo{year}{2000}).


\bibitem[{\citenamefont{Guesten et~al.}(1978)\citenamefont{Guesten, Schuster,
  and Seitz}}]{Guesten1978}
\bibinfo{author}{\bibfnamefont{H.}~\bibnamefont{Guesten}},
  \bibinfo{author}{\bibfnamefont{P.}~\bibnamefont{Schuster}}, \bibnamefont{and}
  \bibinfo{author}{\bibfnamefont{W.}~\bibnamefont{Seitz}},
  \bibinfo{journal}{AIP Conf. Proc.} \textbf{\bibinfo{volume}{82}},
  \bibinfo{pages}{459} (\bibinfo{year}{1978}).

\bibitem[{\citenamefont{Agostinelli et~al.}(2003)}]{Geant}
\bibinfo{author}{\bibfnamefont{S.}~\bibnamefont{Agostinelli}}
  \bibnamefont{et~al.} (\bibinfo{collaboration}{Geant4 Collaboration}),
  \bibinfo{journal}{Nucl. Instr. Meth. A} \textbf{\bibinfo{volume}{506}},
  \bibinfo{pages}{250} (\bibinfo{year}{2003}).

\bibitem[{\citenamefont{Wurm et~al.}(2007)}]{Wurm:2007cy}
\bibinfo{author}{\bibfnamefont{M.}~\bibnamefont{Wurm}} \bibnamefont{et~al.},
  \bibinfo{journal}{Phys. Rev.} \textbf{\bibinfo{volume}{D75}},
  \bibinfo{pages}{023007} (\bibinfo{year}{2007}), 
  \eprint{arXiv:astro-ph/0701305}.

\bibitem[{\citenamefont{Wurm}(2005)}]{MWurm_Dipl}
\bibinfo{author}{\bibfnamefont{M.}~\bibnamefont{Wurm}},
  \bibinfo{note}{diploma thesis, Technische Universit\"at M\"unchen, Garching,
  Germany} (\bibinfo{year}{2005}).


\end{thebibliography}
\end{document}